\documentstyle[prd,aps,floats,psfig]{revtex}
\begin{document}
\preprint{BROWN-HET-1160, IC-98-236, hep-ph/yymmddd}
\draft 

\renewcommand{\topfraction}{0.99}
\renewcommand{\bottomfraction}{0.99}
\twocolumn[\hsize\textwidth\columnwidth\hsize\csname 
@twocolumnfalse\endcsname

\title{\Large Non-thermal Production of Neutralino Cold Dark Matter from 
Cosmic String Decays}

\author{R. Jeannerot$^1$, X. Zhang$^{2,1}$ and R. Brandenberger$^3$}
\address{~\\$^1$International Centre for Theoretical Physics (ICTP), POB 586, 
Strada Costiera 11, 34014 Trieste, Italy;
~\\$^2$CCAST (World Laboratory), P.O. Box 8730, Beijing 100080, P.R. China, 
and Institute of High Energy Physics, Academia Sinica,
Beijing 100039, P.R. China;
~\\$^3$Department of Physics, Brown University, 
Providence, RI 02912, USA}
\date{\today} 
\maketitle

\begin{abstract}
We propose a mechanism of nonthermal production of a neutralino cold dark 
matter particle, $\chi$, from the decay of cosmic strings which form from 
the spontaneous breaking of a $U(1)$ gauge symmetry, such
as $U_{B-L}(1)$, in an extension of the minimal supersymmetric standard model 
(MSSM). By explicit calculation, we point out that with a symmetry breaking 
scale $\eta$ of around $10^8$ GeV, the decay of cosmic strings can give rise to
$\Omega_\chi \simeq 1$. This gives a new constraint on supersymmetric models. 
For example, the dark matter produced from strings will 
overclose the universe if $\eta$ is near the electroweak symmetry breaking 
scale. To be consistent with $\Omega_\chi \leq 1$, the mass of the new 
$U(1)$ gauge boson must be much larger than the Fermi scale which makes it 
unobservable in upcoming accelerator
experiments. In a supersymmetric model with an extra $U_{B-L}(1)$ symmetry, 
the requirement of $\Omega_\chi \leq 1$ puts an
upper bound on the neutrino mass of about $30 eV$ provided neutrino masses are 
generated by the see-saw mechanism.
\end{abstract}

]

\section{Introduction}

In spite of the increasing evidence that cold matter (matter with pressure 
$p = 0$) makes up less than the critical density $\rho_c$ for a spatially flat 
Universe, equally strong evidence for the existence of a substantial amount of 
cold dark matter (CDM) remains. The best current estimates give $\Omega_{CDM} 
\sim 0.3$ \cite{Bahcall} whereas $\Omega_B < 0.1$ \cite{BBN} (here, 
$\Omega_X = \rho_X / \rho_c$ denotes the fractional contribution of $X$ matter 
to $\rho_c$, and $B$ stands for the contribution of baryons).

The leading candidates for cold dark matter are the axion and the neutralino.
The axion is a neutral spin-zero Pseudo-Goldstone boson associated with the 
spontaneous breaking of
the global $U_{PQ}(1)$ symmetry, which was introduced by Peccei and Quinn 
\cite{PQ} as a solution to the strong
CP problem. At zero temperature the axion mass is given by
$$ m_a \sim 6 \times 10^{-6} eV N ( \frac{10^{12}\,\, {\rm  GeV}}{f_a} ) $$
where $f_a$ is Peccei-Quinn symmetry breaking scale and N is a positive 
integer which describes the
color anomaly of $U_{PQ}(1)$. Axions can be produced by three different 
mechanisms: vacuum alignment,
axion string decay and axion domain wall decay \cite{decay}. Cosmology yields 
an upper limit on $f_a$ of $f_a \leq 10^{12}$ GeV.

The neutralino is an electrically neutral hypothetical particle which arises 
in supersymmetric models. In many such models, e.g. in the MSSM (the minimal 
supersymmetric standard model), 
the lightest supersymmetric particle (LSP) is stable, unless R-parity 
violating interactions are
included. The LSP is generally thought to be the lightest neutralino $\chi$. 
The neutralinos in the Universe today are in general assumed to be a relic of 
an initially thermal neutralino distribution in the hot early Universe. Based 
on this thermal production mechanism, there have been many calculations of the 
LSP abundance (for a review, see e.g. \cite{JKG}) as a function of
the MSSM parameters. These studies show that there exists a domain of 
parameter space in the MSSM which is consistent with all of the present 
experimental constraints and for which the
$\chi$ has a relic mass density $\Omega_\chi \sim 1$. However, cosmology also 
imposes limits on the
LSP mass. In the case of a Bino-like LSP, the calculation of Refs.\cite{Shred} 
yields $M_{\tilde B} \leq 300$ GeV. A recent study \cite{Ellis}
relaxes this upper bound to about 600 GeV by including the $\tilde B$ 
coannihilations with the $\tilde e$ and $\tilde \mu$.  

In this paper, we propose a new non-thermal production mechanism of the LSP. 
We consider models with an extra
$U(1)$ gauge symmetry in extensions of the MSSM. This $U(1)$ 
symmetry could be  
$U_{B-L}(1)$, where $B$ and $L$ are respectively baryon and lepton
numbers. Such models
explain the neutrino masses via the see-saw mechanism. Another possibility is 
that the new $U(1)$ corresponds to a $U(1)^\prime$ from string theory 
or grand unified theories  \cite{Langacker}. 

The basic idea of our mechanism is as follows. When the extra $U(1)$ symmetry 
which we have introduced gets broken at a scale $\eta$, a network of 
strings is produced by the usual Kibble mechanism \cite{Kibble}. The initial 
separation of the strings is microscopic, of the order 
$\lambda^{-1} \eta^{-1}$ (where $\lambda$ is a typical Higgs self coupling 
constant of the $U(1)$ sector of the theory) which implies that a substantial 
fraction of the energy density of the Universe is trapped in strings. After 
the symmetry breaking phase transition, the defect network coarsens. In the 
process, string loops decay. If, as we assume, the fields excited in the 
strings couple to the neutralino $\chi$, then a non-thermal distribution of 
$\chi$ particles will be generated during the process of string decay. The 
total energy density in $\chi$ particles will depend on the scale $\eta$ of 
$U(1)$ symmetry breaking. The presence of our alternative generation mechanism 
for $\chi$ particles relaxes the constraints on the mass of the $\chi$. Even 
if the usual thermal generation mechanism is too weak to generate 
$\Omega_{\chi} \sim 1$, our new non-thermal mechanism may, for appropriate 
values of $\eta$, be able to lead to $\Omega_{\chi} \sim 1$. In fact, we find 
that if $\eta < 10^8{\rm GeV}$ and $M_{\chi} \sim 100{\rm GeV}$, then our mechanism will lead to 
$\Omega_{\chi} > 1$, unless the couplings of the $U(1)$ sector to $\chi$ are 
small. Note that there are similarities between our non-thermal production and
the mechanism based on preheating proposed in \cite{Chung}. 

To begin with, we consider a general case and calculate the relic mass density 
of the LSP, then we will move on to a discussion of some implications.   

\section{LSP Production via String Decay}

Local cosmic strings form at a phase transitions associated with the
spontaneous symmetry breaking of a gauge group $G$ down to a subgroup
$H$ of $G$ if the first homotopy group of the vacuum manifold 
$\pi_1({G\over H})$ is nontrivial. We suppose the existence of such a phase 
transition which is
induced by the vacuum expectation value (vev) of some Higgs field $\Phi$, 
$\langle |\Phi |
\rangle = \eta$, and takes place at a
temperature $T_c$ with $T_c \simeq \eta$. The strings are formed by the Higgs 
field $\Phi$
and some gauge field $A$ of $G$ whose generator is broken by the
vev of $\Phi$. We assume that the generator of $G$
associated with $A$ is diagonal so that the strings are abelian. The
mass per unit length of the strings is given by $\mu = \eta^2$. 

During the phase transition, a network of strings forms, consisting of both 
infinite strings and cosmic string loops. After the transition, the infinite 
string network coarsens and more loops form from the
intercommuting of infinite strings. Cosmic string loops loose their energy by 
emitting gravitational
radiation. When the radius of a loop becomes of the order of the string
width, the loop releases its final energy into a burst of $\Phi$ and
$A$ particles. Those particles subsequently decay into LSP, which we denote by 
$\chi$, with
branching ratios $\epsilon$ and $\epsilon'$. For simplicity we now
assume that all the final string energy goes into $\Phi$ particles. A single
decaying cosmic string loop thus releases $N \simeq 2 \pi \lambda^{-1}
\epsilon$ LSPs which we take to have a monochromatic distribution with energy 
$E \sim {T^c \over 2}$.

In such scenarios, we thus have two sources of cold
dark matter which will contribute to the matter density of the universe. 
We have CDM which comes from the standard
scenario of thermal production; it gives a contribution to the
matter density $\Omega_{therm}$. And we also have non-thermal 
production of CDM which comes from the decay of cosmic string loops and gives 
a contribution
$\Omega_{nonth}$. The total CDM density is $\Omega_{CDM} =
\Omega_{therm}+ \Omega_{nonth}$. During the temperature interval between
$T_c$ and the LSP freezeout temperature $T_\chi$, LSPs
released by decaying comic string loops will thermalise very quickly
with the surrounding plasma, and hence will contribute to
$\Omega_{therm}$, which should not sensitively deviate from the value
calculated by the
standard method \cite{JKG,Ellis}.  However, below the LSP freezeout
temperature, since the annihilation of the LSP is by definition
negligible, each CDM particle released by cosmic string decays will
contribute to $\Omega_{nonth}$. We obviously must have 
\begin{equation}
\Omega_{nonth}<1 \label{eq:bound}.
\end{equation}
This will lead us to a constraint (a lower bound) on the cosmic string
forming scale. We now calculate
$\Omega_{nonth}$. 

We assume that the strings evolve in the friction dominated regime so
that the very small scale structure on the strings has not formed
yet. The network of strings can then be described by a single length
scale $\xi(t)$ {\footnote{The friction dominated regime lasts from the
time $t_c$ at which the strings network forms until a time $t_* \sim (G
\mu )^{-1}t_c$, where $G$ is Newton's constant \cite{ShelVil}. In our
scenario, the CDM is produced at and below the LSP freezeout temperature
$T_\chi \sim 10^2-10^3 $ GeV. Hence for $T_c \leq
(10^{10.5}-10^{11}) {\rm GeV} =T_c^*$, when the temperature of the Universe
reaches $T_\chi$, the strings are still in the friction dominated
regime. Since we are looking for a lower bound $T_c^l$ on the scale $\eta$ of 
the strings, and since as we will show below $T_c^l \ll T_c^*$, the time 
interval of interest in our
scenario is in the friction dominated regime.}}. In the
friction dominated period, the length scale $\xi(t)$ has been shown to
scale as \cite{RobRiotto}:
\begin{equation}
\xi(t) = \xi(t_c) \left ({t \over t_c}\right )^{3\over 2} \label{eq:xi}
\end{equation}
where $\xi(t_c) \sim (\lambda \eta )^{-1}$ and $\lambda$ is the Higgs
self quartic coupling constant. The number density of cosmic string loops 
created per unit of time is
   given by \cite{ShelVil}:
\begin{equation}
{dn\over dt} = \nu \xi^{-4} {d\xi \over dt}
\end{equation}
where $\nu$ is a constant of order 1. We are interested in loops
   decaying below $T_\chi$. 

The number density of LSP released from $t_{lsp}$ till today is given by:
\begin{equation}
n^{nonth}_{lsp}(t_0) = N \nu \int^{\xi_0}_{\xi_F} \left ( {t \over t_0} \right
)^{3\over 2} \xi^{-4} d\xi \label{eq:no}
\end{equation} 
where the subscript $0$ refers to parameters which are
evaluated today.
$\xi_F = \xi(t_F)$ where $t_F$ is the time at which cosmic string
loops which are decaying at time $t_{\chi}$ (associated with the LSP freezeout 
temperature $T_{\chi}$) have formed. Now the loop's average radius
shrinks at a rate \cite{ShelVil} ${dR\over dt} = - \Gamma_{loops} G \mu$,
where $\Gamma_{loops}$ is a numerical factor $\sim 10-20$. Since loops
form at time $t_F$ with an average radius $R(t_F) \simeq \lambda^{-1} G \mu
M_{pl}^{1\over 2} t_F^{3\over 2}$, they have shrunk to a point at the time
$t \simeq \lambda^{-1} \Gamma^{-1}_{loops} M_{pl}^{1\over 2} t_F^{3\over 2}$.
Thus 
$t_F \sim (\lambda\Gamma)^{2\over 3}_{loops} M_{pl}^{-{1\over 3}} 
t_\chi^{2\over 3}$.
Now the entropy density is $ s = {2 \pi^2
\over 45} g_* T^3$ where $g_*$ counts the number of massless degrees of
freedom in the corresponding phase. The time $t$ and temperature $T$ are 
related by $t = 0.3 g_*^{-{1\over 2}}(T) {M_{pl}\over T^2}$ where $M_{pl}$ is the 
Planck
mass. Thus using Eqs.(\ref{eq:xi}) and (\ref{eq:no}), we find
that the LSP number density today released by decaying cosmic string
loops is given by:
\begin{equation}
Y^{nonth}_{LSP} = {n^{nonth}_{lsp}\over s} = {{6.75} \over {\pi}} \epsilon \nu
\lambda^2 \Gamma_{loops}^{-2} g_{*_{T_c}}^{-9 \over 4}
g_{*_{T_\chi}}^{3 \over 4} \,  
 M_{pl}^2\, {T_{\chi}^4 \over T_c^6} \, , \label{eq:Ynonth}
\end{equation}
where the subscript on $g^*$ refers to the time when $g^*$ is evaluated.

The LSP relic abundance is related to $Y_{\chi}$ by:
\begin{eqnarray}
\Omega_\chi\, h^2 & \approx & M_{\chi} Y_{\chi} s(t_0) \rho_c(t_0)^{-1} h^2 \nonumber \\      & \approx & 2.82 \times 10^8\, Y^{tot}_\chi\, 
(M_{\chi}/{\rm GeV}) \label{eq:Omega}
\end{eqnarray}
where $h$ is the Hubble parameter and $M_\chi$ is the LSP mass.  
Now $Y^{tot}_{LSP} =
Y^{therm}_\chi+ Y^{nonth}_\chi$; hence by setting $h=0.70$, Eqs. 
(\ref{eq:Omega}) and (\ref{eq:bound}) lead to the following constraint:
\begin{equation}
5.75 \times 10^8\, Y^{nonth}_\chi\, (M_{\chi}/{\rm GeV}) < 1. \label{eq:Yb}
\end{equation}
We thus see that Eqs. (\ref{eq:Ynonth}) and (\ref{eq:Yb}) lead to a lower bound
on the cosmic string forming temperature $T_c$.
 
Recent measurements of cosmological parameters from the cosmic
microwave background radiation combined with Type IA supernovae show
evidence for a cosmological constant. In such a scenario, the
relic matter density satisfies \cite{SN} $\Omega_M h^2 \simeq 0.35$.

\begin{figure}
\psfig{file=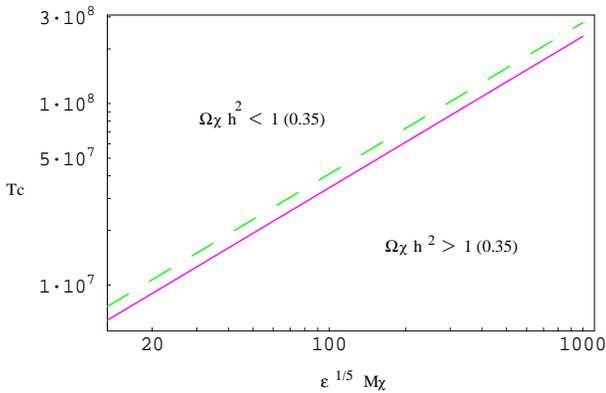,width=8cm}
\caption{The critical temperature $T_c$ as a function of the branching
ratio and the LSP mass
 for $\Omega_\chi h^2 =1$ (solid line) and
$\Omega_\chi h^2 =0.35$ (dashed line). The region above the curves corresponds
to $\Omega_\chi h^2<1$ ($\Omega_\chi h^2<0.35$ respectively) and the region
below correponds to $\Omega_\chi h^2>1$ ($\Omega_\chi h^2>0.35$). The
latter is excluded by observations.}
\label{fig:Omega}
\end{figure}

In Fig. \ref{fig:Omega}, we
have plotted the bound on $T_c$ as a function of $\epsilon^{1\over 5} M_\chi$ 
for both $\Omega_\chi h^2=1$ and  
$\Omega_\chi h^2=0.35$. We have set $g_{*_{T_c}} = 250$,
$g_{*_{T_\chi}}= 90$,  $T_\chi = {m_\chi
\over 20}$, $M_{pl} = 1.22 \times 10^{19}$ GeV, and  the cosmic  
string parameters $\nu =1$, $\lambda = 0.5$ and $\Gamma = 10$.
 The region above each curves corresponds to $\Omega_\chi h^2<1$ ($\Omega_\chi
h^2 < 0.35$ respectively), and the 
region below to $\Omega_\chi h^2 >1$ ($\Omega_\chi h^2 >0.35$ respectively); this 
region is excluded  by observations. We see that if there is a
cosmological constant, a slightly stronger bound on $T_c$ is obtained.

\section{Implications for Phenomenology}

Our results have important implications for supersymmetric extensions of the
standard model with extra $U(1)$'s (or grand unified models with an
intermediate $SU(3)_c \times SU(2)_L \times U(1)_Y \times U(1)'$ gauge
symmetry). Most importantly, the requirement $\Omega_{nonth} < 1$ imposes a new constraint on supersymmetric model building and rules out many models with a low scale of a new symmetry breaking which produces defects such as cosmic strings.
 
Consider, for example, the model with an  
extra $U_{B-L}(1)$ gauge symmetry. In this model,
the spectrum of the standard model is extended to include right-handed
neutrinos 
$N_i$. The light neutrinos receive masses via the see-saw mechanism and the
matter-antimatter asymmetry of the universe is generated by the
out-of-equilibrium decay of these right-handed neutrinos. In the latter case, 
leptogenesis can occur by
the decay of cosmic strings associated with the spontaneous breaking of the
$U_{B-L}(1)$ gauge symmetry \cite{leptrach}. In the  supersymmetric version of 
this model, the strings will release not only right-handed neutrinos
$N_i$, but also their superpartners $\tilde N_i$. The heavy neutrinos $N_i$ and
their scalar partners $\tilde N_i$ can decay into various final states 
including the LSP. The superpotential relevant to the decays is
$$ W = H_1 \epsilon L y_l E^c + H_2 \epsilon L y_\nu N^c ,$$
where $H_1, H_2, L, E^c$ and
$N^c$ are the chiral superfields and $y_l, y_\nu$ are Yukawa couplings for the 
lepton and neutrino Dirac masses, $m_l = y_l v_1, m_D = y_\nu v_2$, with 
$v_{1,2}$ being the
vacuum expectation values of the Higgs fields. At tree level, 
the decay rates of $N_i$ into s-lepton plus Higgsino and lepton plus Higgs are
the same and they are smaller than the rate of $\tilde N_i$ decaying into
s-lepton plus Higgs and
Higssino plus lepton by a factor of 2. If the neutralino is higgsino-like, 
the LSP arise directly from the decays of the $N_i$ and $\tilde N_i$. If the 
neutralino
is bino- or photino-like, subsequent decays of s-lepton into binos or photinos
plus leptons will produce the LSP. For 
reasonable values of the parameters, we estimate the branching ratio $\epsilon$
of the heavy particle decay into LSP to be between 0.1 and 0.5. From Eq. 
(\ref{eq:Ynonth}) it follows that string decays can easily produce
the required amount of LSP. However
too many LSPs will be generated unless the $B-L$ breaking scale,
$\Lambda_{B-L}$ is higher than about  
$10^8$ GeV . In turn, this
will set a lower limit on the neutrino masses generated by the see-saw 
mechanism, $m_\nu \sim m_D^2/ \Lambda_{B-L}$. Inserting numbers and taking 
$m_D \sim m_\tau \sim 1.8$ GeV, one obtains that $m_\nu \leq 30$ eV. 

In models with spontaneous breaking of a $U_{B-L}(1)$ gauge symmetry,
upper and  
lower bounds on the $B-L$ breaking scale have already been derived
from considerations of cosmic rays from string decay \cite{Jane} and
from leptogenesis \cite{BY},
respectively. Our lower bound on the $B-L$ breaking scale is independent of
leptogenesis. 
%especially  the constraints on B-L scale from
%leptogenesis are not valid in the scheme of baryogenesis
%during the electroweak phase transition of SUSY standard model. 

Our lower limit on the $B-L$ symmetry breaking scale in gauged $B-L$ models
and in general models with an extra $U(1)$ \cite{Langacker} pushes the mass of 
the new  gauge boson far above the Fermi scale, 
rendering it impossible to test the new physics signals from the extra 
$Z^\prime$ in accelerators.

To summarize, we have pointed out a new production mechanism for neutralino 
dark matter which can be effective in many models beyond the MSSM, models with 
extra gauge symmetries which admit topological defects. The decay of these 
defects gives rise to a nonthermal contribution to the neutralino density. We 
have focused on the nonthermal production of LSPs from string decays. 
Similarly, one could consider LSP production from other topological defects. 
We have calculated the relic LSP mass density $\Omega_{nonth}$ as a 
function of the string scale, the freezeout temperature and the mass of the LSP. The LSP mass density has two contributions, one from thermal production which 
has been calculated by many authors in the literature before, another is the 
non-thermal production calculated in this paper. Our results indicate that if 
the scale $\eta$ of string production is about $10^8$GeV, then our nonthermal
mechanism can produce the required closure density of LSPs. For values of 
$\eta$ smaller than the above bound, the model is in conflict with 
observations since the LSPs would overclose the Universe. 

One important caveat must be made concerning our calculations. Cosmic strings 
arising in supersymmetric models are generically superconducting \cite{ACD}. 
In this case, the string dynamics may be very different from that of ordinary 
strings, the dynamics assumed in this paper, and thus the corresponding 
constraints on particle physics model building would be quite different. 
Nevertheless, the main point that cosmic string decay in extensions of the 
MSSM can yield a new production mechanism for dark matter remains unaffected.

\centerline{Acknowledgments}

We are grateful to A.-C. Davis and G. Senjanovi\'{c} for useful
discussions. This work is supported
in part by the National Natural Science Foundation 
of China and by the U.S. Department of Energy under 
Contract DE-FG02-91ER40688, TASK A.

\end{document}